\documentclass[aip,jmp,preprint,nofootinbib]{revtex4-1}

\draft 

\usepackage{hyperref}
\usepackage{amsthm,amsfonts,amsmath, amscd}
\usepackage{amssymb,amsmath, dsfont}
\usepackage{ytableau}
\usepackage{physics}
\usepackage{braket}
\usepackage{dsfont}
\usepackage{tikz}

\newtheorem{theorem}{\bf Theorem}[]
\newtheorem{corollary}[theorem]{\bf Corollary}
\newtheorem{proposition}[theorem]{\bf Proposition}
\newtheorem{lemma}[theorem]{\bf Lemma}

\newtheorem{definition}[theorem]{\bf Definition}

\theoremstyle{remark}

\newtheorem*{remarks}{\bf Remarks}

\DeclareMathOperator{\Std}{Std}

\DeclareMathOperator{\SYM}{SYM}
\DeclareMathOperator{\sgn}{sgn}

\begin{document}


\title{Lower bound on entanglement in subspaces defined by Young diagrams}



\author{Robin Reuvers}
\email[Email: ]{r.reuvers@damtp.cam.ac.uk}
\affiliation{$^{1}$Department of Applied Mathematics and Theoretical Physics (DAMTP), Centre for Mathematical Sciences, University of Cambridge, Wilberforce Road, Cambridge CB3 0WA, United Kingdom}

\begin{abstract}
Eigenvalues of 1-particle reduced density matrices of $N$-fermion states are upper bounded by $1/N$, resulting in a lower bound on entanglement entropy. We generalize these bounds to all other subspaces defined by Young diagrams in the Schur--Weyl decomposition of $\otimes^N\mathbb{C}^d$.
\end{abstract}


\maketitle 

\section{Introduction}
The most striking property of fermions is that they satisfy the Pauli exclusion principle: no two can occupy the same state. Mathematically, this is usually formalized by saying that the expectation value of any particle number operator $n_i:=a^\dagger_ia_i$ in a normalized fermionic state $\ket\psi$ is bounded by 1,
\begin{equation}
\label{pnumber}
\braket{n_i}=\bra{\psi}a^\dagger_ia_i\ket{\psi}=\bra{\psi}\mathds{1}-a_ia^\dagger_i\ket{\psi}=1-\|a^\dagger_i\ket\psi\|^2\leq1.
\end{equation}

There is a different way to formulate this. For a normalized $N$-fermion state $\ket\psi$ in the antisymmetric tensor product space $\wedge^N\mathbb{C}^d\subset\otimes^N\mathbb{C}^d$, we can study the \emph{1-particle reduced density matrix}
\begin{equation}
\label{rdm}
\gamma^\psi_1:=\Tr_{1\dots N-1}[\ketbra{\psi}],\ \ \ \ \ \ \ \Tr[\gamma^\psi_1]=1,
\end{equation}
where the trace is over copies $1,\dots,N-1$ of the Hilbert space. (Note though, that the result is the same for any $N-1$ copies because of antisymmetry.)

The equivalent of \eqref{pnumber} is now
\begin{equation}
\label{bound22}
\gamma^\psi_1\leq \frac{1}{N}\mathds{1},
\end{equation}
or that the eigenvalues of $\gamma^\psi_1$ are all bounded by $1/N$. After all, an annihilation operator $a_i$ acts as $\sqrt{N}(\mathds{1}\otimes\bra{i})$ on $N$-fermion states like $\ket\psi$, for some 1-particle state $\ket{i}$, and
\begin{equation}
\label{enumber}
\bra{i}\gamma^\psi_1\ket{i}=\Tr_{1\dots N}[(\mathds{1}\otimes\ketbra{i})\ketbra{\psi}]=\frac{1}{N}\|a_i\ket\psi\|^2=\frac{1}{N}\braket{n_i}\leq\frac{1}{N}.
\end{equation}
A similar bound for bosons gives $N$ in \eqref{pnumber} and $1$ in \eqref{enumber}: no Pauli principle.\\

In this paper, we study $\gamma^\psi_1$ not for bosons or fermions but for other `symmetry types'. These appear alongside the bosonic (fully symmetric) and fermionic (fully antisymmetric) subspaces of $\otimes^N\mathbb{C}^d$ in a decomposition known as Schur--Weyl duality [\onlinecite{GoodmanNolan},\onlinecite{Weyl}],
\begin{equation}
\otimes^N\mathbb{C}^d = \bigoplus_{\nu\vdash N} V^\nu\otimes S^\nu.
\end{equation}
Here, $\nu$ are partitions of $N$---equivalently Young diagrams---and $V^\nu$ are $S^\nu$ are the corresponding irreps of $U(d)$ and $S_N$ respectively. That is, $V^\nu$ encodes the effect of 1-particle basis changes, $\ket\psi\longrightarrow U\otimes\dots\otimes U\ket\psi$, $U\in U(d)$; $S^\nu$ describes what happens under particle permutations, $\ket\psi\longrightarrow U_\sigma\ket\psi$, $\sigma\in S_N$. Schur--Weyl duality says the two are related.

Take $N=3$ with $d\geq3$ as an example, 
\ytableausetup{boxsize=3pt}
\begin{equation}
\otimes^3\mathbb{C}^d = (V^{\ydiagram{3}}\otimes S^{\ydiagram{3}})\oplus (V^{\ydiagram{2,1}}\otimes S^{\ydiagram{2,1}}) \oplus (V^{\ydiagram{1,1,1}}\otimes S^{\ydiagram{1,1,1}}).
\end{equation}
\ytableausetup{boxsize=normal}
Row diagrams correspond to fully symmetric spaces and column diagrams to fully antisymmetric ones, so the space on the left is the bosonic $\otimes^3_{\SYM}\mathbb{C}^d$ and the one on the right the fermionic $\wedge^3\mathbb{C}^d$. It is easy to see with orthogonality that the space in the middle contains
\begin{equation}
\label{som1234}
\ket{\tfrac{1}{\sqrt{2}}(\uparrow\downarrow-\downarrow\uparrow)}\otimes\ket\uparrow,
\end{equation}
where $\ket\uparrow$, $\ket\downarrow$ are any two orthonormal vectors.\\
\ytableausetup{boxsize=3pt}

The question we now ask is: \emph{Is there a bound like \eqref{bound22} for the spaces $V^\nu\otimes S^\nu$?} Equation \eqref{som1234} shows that the trivial bound 1 can be attained for $V^{\ydiagram{2,1}}\otimes S^{\ydiagram{2,1}}$---is that always the case except for fermions? And what states do even appear in these spaces?

The latter was perhaps our main motivation to study this problem: the $V^\nu\otimes S^\nu$ are fundamental objects in representation theory, but little seems to have been published about the entanglement properties of the states within. A bound like \eqref{bound22} is truly the most basic step one can take in this direction---it says that upon Schmidt decomposing a normalized $N$-fermion state $\ket\psi$ as 
\begin{equation}
\ket{\psi}=\sum^d_{i=1} \lambda^\psi_i \ket{\phi_i} \otimes \ket{u_i},
\end{equation}
with $\ket{\phi_i}\in\wedge^{N-1}\mathbb{C}^d$, $\ket{u_i}\in\mathbb{C}^d$, the resulting entanglement entropy is at least $\log(N)$,
\begin{equation}
\label{entbound}
S(\gamma^\psi_1)=-\sum_i(\lambda^\psi_i)^2\ln[(\lambda^\psi_i)^2]\geq -\sum_i(\lambda^\psi_i)^2\ln(\frac{1}{N})=\log(N).
\end{equation}
Here we should emphasize that this is an example of \emph{particle entanglement} [\onlinecite{Schoutens},\onlinecite{entanglement2fermions}]---the entanglement of some of the particles in the system with the remaining ones. We do not discuss \textit{mode entanglement} [\onlinecite{Wolf}]---the entanglement between restrictions of the state to complementary regions of Hilbert space---although that can also be made sense of in this context.\\

Now let us be critical for a moment: if this is \textit{particle} entanglement, which particles are these? Is this still about counting as it was in the case of the Pauli principle \eqref{pnumber}? Although it is useful to note that the spaces $V^\nu\otimes S^\nu$ show up naturally for fermions with spin [\onlinecite{Klyachko}] (where total spin decomposes into irreps of $SU(2)$), in that case there is always an antisymmetrization in the background\footnote{Fermions with spin are described by $\wedge^N(\mathbb{C}^d\otimes\mathbb{C}^2)$, which is embedded in $(\otimes^N\mathbb{C}^d)\otimes(\otimes^N\mathbb{C}^2)$. If we trace out the spin part, the reduced state should still be permutation invariant. Indeed, for fermionic states with definite total spin described by a two-row Young diagram $\nu$ with $N$ boxes, the reduced state is $(\rho^{\nu^t}\otimes\mathds{1}_{S^{\nu^t}})/\dim(S^{\nu^t})$, where $\nu^t$ denotes the transpose and $\rho^{\nu^t}$ is some density matrix on $V^{\nu^t}$. The 1-body reduced density matrices of such states have been completely understood [\onlinecite{Klyachko}], but this is different from the problem studied here.}. In contrast, the particles we are talking about ought not to be indistinguishable, nor should the reduced density matrices all be equivalent like in \eqref{rdm}. So although the notion of creation and annihilation operators can be given meaning with the formalism discussed in Section \ref{reviewour}, it is perhaps best to say that there are no actual particles to be counted. The justification of our question rather lies in the ubiquity of $\otimes^N\mathbb{C}^d$ and Schur--Weyl duality in spin systems [\onlinecite{HeisenbergYoung}] and quantum information theory [\onlinecite{BaconChuangHarrow},\onlinecite{recoupcoeff},\onlinecite{Hayashi},\onlinecite{KeylWerner}]. It is the desire to simply understand $V^\nu\otimes S^\nu$, and to see how it holds up as an explicit example providing insight into entanglement-related questions [\onlinecite{Grudka}].\\

The paper is organized as follows. Section \ref{secmainresult} contains some notation and discusses the main result. Section \ref{somenotation} introduces some more notation. Section \ref{reviewour} is a review of useful past results; we expect it to be of general interest. Section \ref{proof} provides proofs.

\section{Main result}
\label{secmainresult}
The following notation will be used throughout the paper. A \emph{partition of $N$} is denoted $\nu=(r_1,\dots,r_{c_1})$ with integers $r_1\geq r_2\geq\dots\geq r_{c_1}>0$ and $r_1+\dots+r_{c_1}=N$. It defines a \emph{Young diagram} (also called $\nu$) with $c_1$ rows of length $r_1,\dots,r_{c_1}$. This fully determines the length of its $r_1$ columns, which we denote $c_1\geq\dots\geq c_{r_1}$.
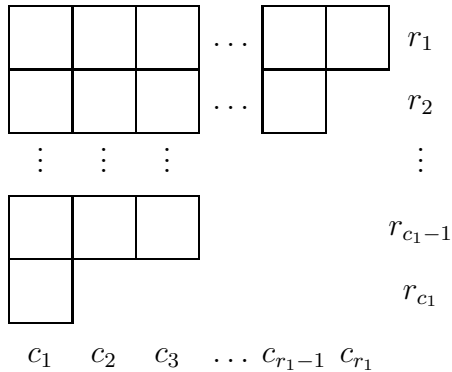
\[
\ytableausetup{mathmode, boxsize=2em}
\begin{ytableau}
 \phantom{.} &  &  & \none[\dots] & & &\none[r_1]\\
\phantom{x} &  &  & \none[\dots] & &\none & \none[r_2] \\
\none[\vdots] & \none[\vdots]
& \none[\vdots] &\none & \none & \none & \none[\vdots]\\
 &  &  & \none &\none&\none&\none[r_{c_1-1}]\\
\phantom{.} & \none & \none &\none&\none& \none&\none[r_{c_1}]\\
\none[c_1] &\none[c_2]&\none[c_3]&\none[\dots]&\none[c_{r_1-1}]&\none[c_{r_1}]
\end{ytableau}
\ytableausetup{boxsize=normal}
\]
We refer to the box in row $i$ and column $j$ as box $(i,j)$. An important quantity is the \emph{hook length} of a box, defined as
\begin{equation}
h_{(i,j)}:=r_i-j+c_j-i+1.
\end{equation}
\emph{Removable boxes} are those that yield a valid Young diagram upon removal. They are positioned at the end of their row \emph{and} column. For example,
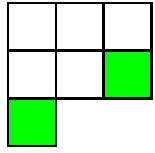
\[
\begin{ytableau}
 \phantom{.}&   &  \\
 \phantom{.}&  & *(green) \\
 *(green)
\end{ytableau}
\]
has two removable boxes, namely $(2,3)$ and $(3,1)$, indicated in green.

\begin{theorem}
\label{mainresult}
Let $\nu$ be a Young diagram of $N\geq 2$ boxes and $r_1$ columns of length $c_1,\dots,c_{r_1}$. Let $d\geq c_1$ and consider
\begin{equation}
\label{setup}
\ket\psi\in V^\nu\otimes S^\nu \subset \bigoplus_{\mu\vdash N} V^\mu\otimes S^\mu = \otimes^N\mathbb{C}^d=(\otimes^{N-1}\mathbb{C}^d)\otimes\mathbb{C}^d.
\end{equation}
Consider the Schmidt decomposition across the final tensor product above, that is, 
\begin{equation}
\label{thisSchmidt}
\ket{\psi}=\sum^{d}_{i=1} \lambda^\psi_i \ket{\phi_i} \otimes \ket{u_i},
\end{equation}
with $\ket{\phi_i}\in\otimes^{N-1}\mathbb{C}^d$, $\ket{u_i}\in\mathbb{C}^d$ and $\lambda^\psi_1\geq\lambda^\psi_2\geq\dots\geq0$. Let $h_{(i,j)}$ denote the hook length of the box $(i,j)$. Then,
\begin{equation}
\label{result}
\sup_{\substack{\ket\psi\in V^\nu\otimes S^\nu\\\|\ket\psi\|=1}}(\lambda^\psi_1)^2=\max_{(c_l,l)\ \textnormal{removable}}\ \prod^{c_{l}-1}_{i=1}\left(1-\frac{1}{h_{(i,l)}}\right).
\end{equation} 
\end{theorem}

\begin{remarks}
\begin{enumerate}
\item For $N$ bosons, only $(1,N)$ is removable and the bound is 1 (from the empty product). For $N$ fermions, only $(N,1)$ is removable and the bound is $1/N$. An easy lower bound for \eqref{result} is the fermionic bound for the shortest column, i.e.\ $1/c_{r_1}$. 
\item The product in \eqref{result} is taken over the boxes above the removable box. For example, the diagram $\nu=(3,2,1)$ has three removable boxes (in green) and the product concerns the boxes in yellow.
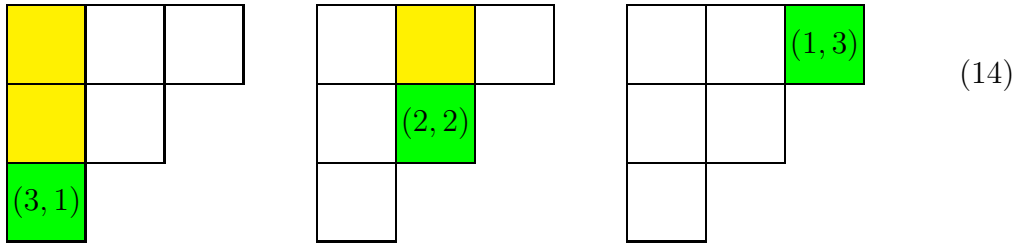
\begin{equation}
\label{examps}
\ytableausetup{mathmode, boxsize=2.5em}
\begin{ytableau}
*(yellow) &   &  \\
*(yellow)  &   \\
*(green) (3,1)
\end{ytableau}\ \ \ \ \ \ \ 
\begin{ytableau}
  \phantom{.}& *(yellow) & \\
  \phantom{.} & *(green) (2,2)   \\
 \phantom{.}
\end{ytableau}
\ \ \ \ \ \ \ 
\begin{ytableau}
  \phantom{.}&  & *(green) (1,3)\\
  \phantom{.} &   \\
 \phantom{.}
\end{ytableau}
\ytableausetup{boxsize=normal}
\end{equation}
The respective bounds are $8/15$, $2/3$ and $1$, so the maximum is $1$. 

\item Each removable box of the diagram $\nu$ gives a bound in \eqref{result}. Sections \ref{reviewour} and \ref{proof} construct interesting subspaces of $V^\nu\otimes S^\nu$ in which this bound holds and is sharp.

\item There is no loss of generality in singling out the final $\mathbb{C}^d$ in \eqref{setup}: if we want to split off copy $1\leq k\leq N-1$ instead, a maximizer is found by swapping spaces $k$ and $N$ in any maximizer for \eqref{result}. Note that this does not say that all cuts are equivalent, and all Schmidt decompositions equal, for a fixed $\ket\psi$---that is only true for bosons and fermions.

\item One could ask if it is necessary to take the maximum over removable boxes in \eqref{result}. Will the box to the right always give the largest product as it did in \eqref{examps}? In reality, there is no easy rule. Below are three examples, with the bounds inserted in the removable boxes and the highest highlighted in green.
\begin{equation}
\ytableausetup{mathmode, boxsize=1.6em}
\begin{ytableau}
 \phantom{.}&   \\
 \phantom{.}&   \\
   \phantom{.}&  \frac13 \\
 *(green) \frac25
\end{ytableau}\hspace{2cm} 
\begin{ytableau}
  \phantom{.}&  & \\
  \phantom{.}&  & \\
  \phantom{.}&  & \frac13\\
  \phantom{.} & *(green)  \frac25 \\
 \frac{8}{21}
\end{ytableau}\hspace{2cm} 
\begin{ytableau}
  \phantom{.}&  & \\
  \phantom{.}&  & \frac12  \\
 *(green) \frac35
\end{ytableau}
\ytableausetup{boxsize=normal}
\end{equation}
\item Our proof in Section \ref{proof} relies on three useful past results about a convenient basis of $S^\nu$, and a reformulation of the maximization problem. These should be of broader interest and are discussed in Section \ref{reviewour}.
\item We construct a maximizer in Section \ref{sharpn}, but are unable to provide necessary and sufficient conditions for the bound to be attained. 
\end{enumerate}
\end{remarks}

As in \eqref{entbound}, the theorem gives a lower bound on (particle) entanglement entropy. 
\begin{corollary}
Consider the set-up of Theorem \ref{mainresult}. Let $\ket\psi\in V^\nu\otimes S^\nu$ be normalized. The entanglement entropy satisfies
\begin{equation}
S(\Tr_{1\dots N-1}[\dyad{\psi}])\geq \min_{(c_l,l)\ \textnormal{removable}}\ \ln\left(\prod^{c_{l}-1}_{i=1}\frac{h_{(i,l)}}{h_{(i,l)}-1}\right).
\end{equation}
\end{corollary}
For fermions, this bound is attained by Slater determinants \eqref{Slater}, but it will not be sharp in general. Improvements or bounds concerning higher-order reduced density matrices can be found with our algorithm [\onlinecite{RR}], employing Theorems \ref{recursion} and \ref{azw} for the required projections [\onlinecite{AlcockZeilingerWeigert},\onlinecite{KeppelerSjodahl}], but given the many open problems for fermions [\onlinecite{CLR},\onlinecite{RR}], we expect that almost all conjectures will be hard to prove.

\section{Some more notation}
\label{somenotation}
It will be convenient to have a general form for the \textit{Schmidt decomposition}: a normalized state $\ket{\psi}$ in a bipartite Hilbert space $\mathcal{H}_A\otimes\mathcal{H}_B$ with dimensions $d_A, d_B$ and $n_s:=\min(d_A,d_B)$ can be written as
\begin{equation}
\label{Schmidt}
\ket{\psi}=\sum^{n_s}_{i=1} \lambda^\psi_i \ket{\psi^i_A} \otimes \ket{\psi^i_B},
\end{equation}
with \emph{Schmidt coefficients} $\lambda^\psi_1\geq\dots\geq\lambda^\psi_{n_s}\geq0$ satisfying $\sum_i(\lambda^\psi_i)^2=1$ and \emph{Schmidt vectors} $\ket{\psi^i_A}\in\mathcal{H}_A$, $\ket{\psi^i_B}\in\mathcal{H}_B$.

We again state the Schur--Weyl duality
\begin{equation}
\label{SchurWeyl}
\otimes^N\mathbb{C}^d = \bigoplus_{\nu\vdash N} V^\nu\otimes S^\nu.
\end{equation}
Using the notation from Section \ref{secmainresult}, these spaces have dimension
\begin{equation}
\label{dimV}
\dim(V^\nu)=\frac{\Pi^{c_1}_{i=1}\Pi^{r_j}_{j=1}(d+j-i)}{\Pi^{c_1}_{i=1}\Pi^{r_j}_{j=1}\ h_{(i,j)}},
\end{equation}
and
\begin{equation}
\dim(S^\nu)=\frac{N!}{\Pi^{c_1}_{i=1}\Pi^{r_j}_{j=1}\ h_{(i,j)}}.
\end{equation}
Note that only the latter depends on the dimension $d$: if we embed $\mathbb{C}^d$ in a larger space, the irrep $V^\nu$ of $U(d)$ gets embedded similarly in its larger equivalent, but the $S_N$-irrep $S^\nu$ remains unchanged.

To build a basis of $S^\nu$, we use \textit{standard Young tableaux}---fillings of the Young diagram with numbers $1,\dots,N$ that increase from left to right and from top to bottom. Denote the set of such diagrams by $\Std(\nu)$ and let $\Std_N:=\cup_{\nu\vdash N}\Std(\nu)$. Given a tableau $t$ we denote the corresponding diagram by $\nu(t)$.

It may be good to mention that we do not distinguish the two relevant tensor product structures---$\mathcal{H}_A\otimes\mathcal{H}_B$ in \eqref{Schmidt} and $V^\nu\otimes S^\nu$ in \eqref{SchurWeyl}---in our notation. It can happen that we act with an operator $B:\mathcal{H}_B\to\mathcal{H}_B$ on a product of $\ket{v}\in V^\nu$ and  $\ket{t}\in S^\nu$, resulting in $(\mathds{1}\otimes B)\ket{v\otimes t}$, but we trust the reader interprets this in the correct way.

Finally, we mention a special fermionic state. An $N$-fermion \emph{Slater determinant} built from orthonormal $\ket{u_1},\dots,\ket{u_N}\in\mathbb{C}^d$ is defined as
\begin{equation}
\label{Slater}
\ket{u_1\wedge\dots\wedge u_N}:=\frac{1}{\sqrt{N!}}\sum_{\sigma\in S_N}(-1)^{\sgn(\sigma)} \ket{u_{\sigma(1)}\otimes\dots\otimes u_{\sigma(N)}}.
\end{equation}

\section{Review of useful results}
\label{reviewour}
The following sections discuss the tools needed in the proof. None of the theorems are new, but they are simple and practical and should be of broad interest.

\subsection{Young's orthogonal basis}
\label{yob}
To get started on our problem, we need a basis of $S^\nu$. This section lists three theorems that provide us with a convenient one. The goal is to obtain orthogonal projections $P_t$ for each standard tableau $t$. We use these to define normalized vectors $\ket{t}\in S^\nu$, but we first need to define some other operators corresponding to $t$.

\begin{definition}
For a standard tableau $t$, let $R_i$ be set of numbers in row $i$ and $C_j$ be the set of numbers in column $j$. Note that $|R_i|=r_i$ and $|C_j|=c_j$. We define symmetric row projections
\begin{equation}
P^S_{i}:=\frac{1}{r_i!}\sum_{\sigma\in \SYM(R_i)}U_\sigma,
\end{equation}
and antisymmetric column projections
\begin{equation}
P^A_j:=\frac{1}{c_j!}\sum_{\sigma\in \SYM(C_j)}(-1)^{\sgn(\sigma)}U_\sigma,
\end{equation}
where $U_\sigma$ is the unitary that implements the permutation $\sigma$ in $\otimes^N\mathbb{C}^d$. 
Also define the row symmetrizer of $t$
\begin{equation}
S_t:=\prod^{c_1}_{i=1} P^S_{i},
\end{equation}
and the column antisymmetrizer
\begin{equation}
A_t:=\prod^{r_1}_{j=1} P^A_j.
\end{equation}
The Young projection of $t$ is defined as 
\begin{equation}
\label{Youngprojection}
Y_t:=\frac{r_1!\dots r_{c_1}! c_1! \dots c_{r_1}!}{\Pi^{c_1}_{i=1}\Pi^{r_j}_{j=1}\ h_{(i,j)}}S_tA_t.
\end{equation}
\end{definition}

The $P^S_i$, $P^A_j$, $S_t$ and $A_t$ are all orthogonal projections, but the $Y_t$ are merely projections in that $(Y_t)^2=Y_t$, but possibly $Y^\dagger_t\neq Y_t$. Often, the basis element of $S^\nu$ corresponding to $t$ is defined through the image of $Y_t$, but since these operators are not hermitian and there  exist $t$ and $s$ such that $Y_tY_s\neq\delta_{ts}Y_t$, this is not the basis we prefer to work in.

Instead, we use \emph{Young's orthogonal basis} (or \emph{form}). It is more convenient for our purposes and probably for doing quantum mechanics in general.
The following theorem defines the basis vectors through its associated projections.

\begin{theorem}[{Thrall [\onlinecite{Thrall}] and Keppeler-Sj\"odahl [\onlinecite{KeppelerSjodahl}]}]
\label{recursion}
Let $N\geq 1$ be an integer. Let $t\in \Std_N$ and let $t^{\downarrow}$ be the tableau with the box containing $N$ removed. For $N\leq2$, let $P_t:=Y_t$. For $N\geq3$, we recursively define
\begin{equation}
\label{defproj}
P_t:=(P_{t^{\downarrow}}\otimes\mathds{1})Y_t (P_{t^{\downarrow}}\otimes\mathds{1}),
\end{equation}
where $Y_t$ is defined in \eqref{Youngprojection} and $P_{t^{\downarrow}}$ acts on $\otimes^{N-1}\mathbb{C}^d$. These operators satisfy
\begin{enumerate}
\item $P_t^2=P_t$, $P_t^\dagger=P_t$, and $P_t=\mathds{1}_{V^{\nu(t)}}\otimes \ketbra{t}$ for normalized vectors $\ket{t}\in S^\nu$,
\item $P_tP_s=\delta_{ts}P_t$,
\item $\sum_{t\in\Std(\nu)}P_t=\mathds{1}_{V^\nu\otimes S^\nu}$ and $\sum_{t\in\Std_N}P_t=\mathds{1}_{\otimes^N\mathbb{C}^d}$.
\end{enumerate}
\end{theorem}

There are two special tableaux for which \eqref{defproj} can be simplified: a tableau is \textit{row-ordered} if we put in the numbers $1,\dots,N$ from left to right, filling up the rows one by one starting from the top; it is \textit{column-ordered} if we put in the numbers from top to bottom, filling up the columns one by one from the left. For example, 
\begin{equation}
\begin{ytableau} 1 & 2 \\ 3 & 4 \\ 5\end{ytableau}\hspace{2cm}\begin{ytableau} 1 & 4 \\ 2 & 5 \\ 3 \end{ytableau}
\end{equation}
are row-ordered and column-ordered respectively.

\begin{theorem}[{e.g.\ Okounkov [\onlinecite{Okounkov}], Alcock-Zeilinger--Weigert [\onlinecite{AlcockZeilingerWeigert}]}]
\label{azw}
A row-ordered tableau $t$ satisfies
\begin{equation}
\label{roword}
P_t=\frac{r_1!\dots r_{c_1}! c_1! \dots c_{r_1}!}{\Pi^{c_1}_{i=1}\Pi^{r_j}_{j=1}\ h_{(i,j)}}S_tA_tS_t.
\end{equation}
A column-ordered tableau satisfies
\begin{equation}
\label{colord}
P_t=\frac{r_1!\dots r_{c_1}! c_1! \dots c_{r_1}!}{\Pi^{c_1}_{i=1}\Pi^{r_j}_{j=1}\ h_{(i,j)}}A_tS_tA_t.
\end{equation}
\end{theorem}

We end this section with the action of $S_N$ on $\ket{t}$, defined through its generators $(k\ k+1)$. 
\begin{theorem}[{Young [\onlinecite{OkounkovVershik},\onlinecite{Thrall},\onlinecite{Young}]}]
\label{Youngrule}
Let $t\in\Std_N$, with $k$ in box $(i,j)$ and $k+1$ in box $(i',j')$. Let $v\in V^{\nu(t)}$. Define $r=(j'-i')-(j-i)$ and let $\tilde{t}$ be the tableau with $k$ and $k+1$ interchanged. The action of the transposition $(k\ k+1)$ on $\ket{v\otimes t}$ is
\begin{equation}
U_{(k\ k+1)}\ \ket{v\otimes t} = \begin{cases}
 \ \ \ \  \ket{v\otimes t}  & \text{if $k$ and $k+1$ are in the same row} \\
 \ \ -\ket{v\otimes t} & \text{if $k$ and $k+1$ are in the same column}\\
\frac{1}{r}\ket{v\otimes t}+\sqrt{1-\frac{1}{r^2}}\ket{v\otimes \tilde{t}} & \text{otherwise}
  \end{cases}.
\end{equation}
\end{theorem}
\ytableausetup{boxsize=3pt}
For example, for $\ket{v}\in V^{\ydiagram{2,1}}$,
\ytableausetup{boxsize=12pt}
\begin{equation}
U_{(23)}\ \big|v\otimes\begin{ytableau} 1 & 2 \\ 3 \end{ytableau}\ \big\rangle= -\frac12\ \big|v\otimes\begin{ytableau} 1 & 2 \\ 3 \end{ytableau}\ \big\rangle\ +\ \sqrt{\frac34}\ \big|v\otimes\begin{ytableau} 1 & 3 \\ 2 \end{ytableau}\ \big\rangle\ .
\end{equation}
\ytableausetup{boxsize=normal}

\subsection{Removing boxes and the Schmidt decomposition}
\label{removebox}
\begin{figure}
\[
\tikz {
\node (a) at (5,0) [circle] {\begin{ytableau} *(green) 1\end{ytableau}};
\node (b) at (3.5,-2.5) [circle] {\ydiagram{2}};
\node (c) at (6.5,-2.5) [circle] {\begin{ytableau} 1\\*(green) 2\end{ytableau}};
\node (d) at (2,-5) [circle] {\ydiagram{3}};
\node (e) at (5,-5) [circle] {\begin{ytableau} 1 & *(green) 3\\ 2\end{ytableau}};
\node (f) at (8,-5) [circle] {\ydiagram{1,1,1}};
\node (g) at (-1,-8.5) [circle] {\ydiagram{4}};
\node (h) at (2,-8.5) [circle] {\ydiagram{3,1}};
\node (i) at (5,-8.5) [circle] {\ydiagram{2,2}};
\node (j) at (8,-8.5) [circle] {\begin{ytableau} 1 & 3\\ 2\\ *(green) 4\end{ytableau}};
\node (k) at (11,-8.5) [circle] {\ydiagram{1,1,1,1}};
\draw (a) edge[->] (b) (a) edge[<->, ultra thick] (c) (b) edge[->] (d) (b) edge[->] (e) (c) edge[<->, ultra thick] (e) (c) edge[->] (f) (d) edge[->] (g) (d) edge[->] (h) (e) edge[->] (h) (e) edge[->] (i) (e) edge[<->, ultra thick] (j) (f) edge[->] (j) (f) edge[->] (k);
}
\]
\caption{Each element of Young's orthogonal basis can be thought of as a path down Young's lattice. The symmetries of the previous tableaux are preserved, making this a convenient basis for Schmidt decompositions.}
\label{tree}
\end{figure}
Theorem \ref{Youngrule} reveals an important aspect of Young's orthogonal basis: $S_{N-1}$ acts on tableaux based only on the position of $1,\dots,N-1$. Hence, building up tableaux by adding boxes preserves the action of the subgroups $S_1\subset S_2\subset\dots$, and splitting off the last space results in Schmidt vectors corresponding to the previous tableau. Or, as we prove below, \emph{to remove a particle is to remove a box}. We include Figure \ref{tree} below to guide the reader's imagination.\footnote{
Young's orthogonal basis is an example of a more general construction named after Gelfand--Zetlin [\onlinecite{GZ1},\onlinecite{GZ2},\onlinecite{OkounkovVershik}]. It hinges on the fact that the restriction of an irrep of $S_N$ to the group $S_{N-1}$ has a multiplicity-free decomposition into irreps of that group, resulting in the lattice of Figure \ref{tree}. The framework can also be applied to the unitary and orthogonal groups [\onlinecite{GZ1}\onlinecite{GZ2},\onlinecite{OkounkovVershik}]. Ignoring this, we use only theorems from Section \ref{yob}.}

\begin{corollary}[Removing one box]
\label{sympres}
Let $t\in\Std_N$. Erase the box with $N$ from $t$ and call the resulting tableau $t^\downarrow$. Let $\ket{v}\in V^{\nu(t)}$ and consider the Schmidt decomposition \eqref{thisSchmidt} of $\ket{v\otimes t}$, with Schmidt vectors $\{\ket{\phi_i}\}$, $\{\ket{u_i}\}$. Then, $\ket{\phi_i}\in \Im(P_{t^\downarrow})$.
\end{corollary}
\begin{proof}
We have $\ket{v\otimes t}=P_t\ket{v\otimes t}$. By Theorem \ref{recursion}, 
\begin{equation}
\lambda^{v\otimes t}_i\ket{\phi_i}=(\mathds{1}\otimes \bra{u_i})\ket{v\otimes t}=P_{t^{\downarrow}}(\mathds{1}\otimes \bra{u_i})Y_t (P_{t^{\downarrow}}\otimes\mathds{1})\ket{v\otimes t}\in \Im(P_{t^\downarrow}).
\end{equation}
\end{proof}

\ytableausetup{boxsize=12pt}
In other words, vectors corresponding to a definite tableau have Schmidt vectors that correspond to the tableau with the box containing $N$ erased---to remove a particle is to remove a box.

Does this generalize to Schmidt decompositions \eqref{Schmidt} with $\mathcal{H}_A=\otimes^{k}\mathbb{C}^d$ and $\mathcal{H}_B=\otimes^{N-k}\mathbb{C}^d$? It does for the Schmidt vectors $\ket{\psi^A_i}$ by the argument above and repeated use of Theorem \ref{recursion}. Alternatively: the action of $S_{k}$ described by Theorem \ref{Youngrule} is independent from $k+1,\dots N$, and for $\mathcal{H}_A$ this corresponds to the tableau with those numbers removed. The situation is more complicated for $\mathcal{H}_B$ as we may not end up with a standard tableau. For example for $k=1$, $\ytableaushort{ 1 3, 2}*{2,1}$ results in $\begin{ytableau} \none & 3 \\ 2 \end{ytableau}$ which we cannot identify with a tableau. However, for $k=2$, $\ytableaushort{ 1 3, 2 4}*{2,2}$ results in $\ytableaushort{3,4}*{1,1}$, which simply tells us that the Schmidt vectors $\ket{\psi^B_i}$ are antisymmetric. The following statement formalizes this and generalizes Corollary \ref{sympres}.
\ytableausetup{boxsize=normal}
\begin{corollary}[Schmidt decompositions]
\label{split}
Let $t\in\Std_N$, $\ket{v}\in V^{\nu(t)}$ and $1\leq k\leq N-1$. Let $t_A$ be the standard tableau obtained by erasing the boxes containing $k+1,\dots, N$. Then,
\begin{equation}
(P_{t_A}\otimes \mathds{1}_{\otimes^{N-k}\mathbb{C}^d})\ket{v\otimes t}=\ket{v\otimes t},
\end{equation}
and the Schmidt vectors in \eqref{Schmidt} with $\mathcal{H}_A=\otimes^{k}\mathbb{C}^d$ and $\mathcal{H}_B=\otimes^{N-k}\mathbb{C}^d$ satisfy $\ket{\psi^A_i}\in \Im(P_{t_A})$.

Now assume that erasing the boxes containing $1,\dots, k$ from $t$ and subtracting $k$ from all entries also results in a standard tableau $t_B$. We then have 
\begin{equation}
(\mathds{1}_{\otimes^{k}\mathbb{C}^d}\otimes P_{t_B})\ket{v\otimes t}=\ket{v\otimes t},
\end{equation}
and $\ket{\psi^B_i}\in \Im(P_{t_B})$. We conclude in this case, that
\begin{equation}
\Im(P_{t})\subset \Im(P_{t_A})\otimes\Im(P_{t_B}).
\end{equation}
\end{corollary}
\begin{proof}
The projection $P_{t_A}$ is defined in terms of elements of the symmetric group. Its action is determined by Theorem \ref{Youngrule}, and is not influenced by $k+1,\dots,N$. We then know that $P_{t_A}\otimes \mathds{1}_{\otimes^{N-k}\mathbb{C}^d}$ leaves $\ket{t}$ invariant since $P_{t_A}$ leaves $\ket{t_A}$ invariant by Theorem \ref{recursion}. The same argument holds for $P_{t_B}$ under the assumptions made. The statement about the Schmidt vectors follows by taking inner products $(\mathds{1}\otimes\bra{\psi^i_B})\ket{v\otimes t}$ and $(\bra{\psi^i_A}\otimes\mathds{1})\ket{v\otimes t}$.
\end{proof}

\subsection{Maximal eigenvalues of reduced density matrices}
\label{fermi}
Here, we discuss a reformulation of the maximization problem \eqref{result}. It relies on a basic variational characterization that is an example of similar, more general statements [\onlinecite{Cao},\onlinecite{RR}],
\begin{equation}
\label{varchar}
\lambda^\psi_1=\sup_{\substack{\ket\alpha\in\mathcal{H}_A,\ \ket\beta\in\mathcal{H}_B\\\|\ket\alpha\|=1,\ \|\ket\beta\|=1}}|\braket{\alpha\otimes\beta|\psi}|.
\end{equation}

\begin{theorem}[{Coleman [\onlinecite{Coleman}], can be generalized to sums [\onlinecite{RR}]}]
\label{normlemma}
Consider the Schmidt decomposition \eqref{Schmidt}. Let $U\subset\mathcal{H}$ be a subspace and let $P_U$ be the orthogonal projection onto $U$. Then,
\begin{equation}
\label{normstate}
\sup_{\substack{\ket\psi\in U\\\|\ket\psi\|=1}} (\lambda^\psi_1)^2 = \sup_{\substack{\ket\alpha\in\mathcal{H}_A,\ \ket\beta\in\mathcal{H}_B\\\|\ket\alpha\|=1,\ \|\ket\beta\|=1}}\|P_U(\ket{\alpha\otimes\beta})\|^2.
\end{equation}
A maximizing $\ket\psi$ satisfies 
\begin{equation}
\label{fixedpoint}
\ket\psi=\frac{P_U\ket{\psi^1_A \otimes \psi^1_B}}{\|P_U\ket{\psi^1_A \otimes \psi^1_B}\|}.
\end{equation}
\end{theorem}
\begin{proof}
One side of the inequality follows from
\begin{equation}
\label{someineq1}
(\lambda^\psi_1)^2=|\bra{\psi}P_U\ket{\psi^1_A \otimes \psi^1_B}|^2\leq\|P_U\ket{\psi^1_A \otimes \psi^1_B}\|^2,
\end{equation}
where the orthogonal projection $P_U$ was placed in with $P^\dagger_U=P_U$ and $P_U\ket{\psi}=\ket{\psi}$ and we used the Cauchy--Schwarz inequality. 
Also, by \eqref{varchar}, for any $\ket\alpha\in\mathcal{H}_A$ and $\ket\beta\in\mathcal{H}_B$,
\begin{equation}
\label{someineq2}
\|P_U\ket{\alpha \otimes \beta}\|=\bra{\alpha \otimes \beta}\frac{P_U\ket{\alpha \otimes \beta}}{\|P_U\ket{\alpha \otimes \beta}\|}\leq \lambda_1^{\frac{P_U\ket{\alpha \otimes \beta}}{\|P_U\ket{\alpha \otimes \beta}\|}},
\end{equation}
which proves \eqref{normstate}. A maximizing $\ket\psi$ has to have equality in \eqref{someineq1} since otherwise \eqref{someineq1} and \eqref{someineq2} contradict the fact that it is a maximizer, which proves \eqref{fixedpoint}.
\end{proof}

We can check what this gives for the basic fermionic bound \eqref{bound22}. If $P^A$ is the projection onto $\wedge^k\mathbb{C}^d$, we conclude that for any $\ket\phi\in\otimes^{k-1}\mathbb{C}^d$ and $\ket{u}\in\mathbb{C}^d$,
\begin{equation}
\label{estim}
\|P^A(\ket\phi\otimes\ket{u})\|^2\leq \frac{1}{k}\|\ket\phi\|^2\|\ket{u}\|^2.
\end{equation}
Coleman [\onlinecite{Coleman}] showed that this is attained if and only if $(\mathds{1}\otimes\bra{u})\ket{\phi}=0$: expanding in Slater determinants in a basis containing $\ket{u}$ easily shows that the condition is sufficient. It is also necessary since \eqref{fixedpoint} implies that 
\begin{equation}
(\mathds{1}\otimes\bra{u})\ket{\phi}=(\mathds{1}\otimes\bra{u}\otimes\bra{u})\frac{P^A\ket{\phi \otimes u}}{\|P^A\ket{\phi \otimes u}\|^2}=0
\end{equation}
as $\bra{u}\otimes\bra{u}$ is symmetric.

\section{Proof of Theorem \ref{mainresult}}
\label{proof}
The main steps in the proof are three reductions and one estimate.
\begin{enumerate}
\item Reduce to the spaces $\Im(P_t)$ using the orthonormal structure from Section \ref{removebox}---that is, work with definite tableaux $t$ rather than linear combinations (Lemma \ref{firstlem}).
\item Reduce to tableaux with $N$ in the various removable boxes and $1,\dots,N-1$ in column-order using the same techniques (Lemma \ref{indep}).
\item Reduce to the case where $N$ is in the first column using Corollary \ref{split} and the column-order (equation \eqref{ineq1}).
\item Prove the bound using the projection-based formulation from Section \ref{fermi} and the projections $P_t$ from Section \ref{yob} (Lemma \ref{firstbox}).
\item Construct states that attain the bound (Section \ref{sharpn}).
\end{enumerate}
\subsection{Upper bound}
Our first reduction step is the following lemma. It relies on the fact that the vectors $\ket{t}$ are orthogonal, and that this remains the case after erasing the box with $N$, so that linear combinations are irrelevant for our maximization problem.
\begin{lemma}
\label{firstlem}
\begin{equation}
\sup_{\substack{\ket\psi\in V^\nu\otimes S^\nu\\\|\ket\psi\|=1}}(\lambda^\psi_1)^2=\max_{t\in\Std(\nu)}\sup_{\substack{\ket{v}\in V^\nu\\\|\ket{v}\|=1}}(\lambda^{v\otimes t}_1)^2
\end{equation}
\end{lemma}
\begin{proof}
We note that for normalized $\ket\psi\in V^\nu\otimes S^\nu$, $\ket\phi\in\otimes^{N-1}\mathbb{C}^d$, $\ket u\in\mathbb{C}^d$, by Cauchy--Schwarz,
\begin{equation}
|\braket{\phi\otimes u|\psi}|^2=|\bra{\phi}(\mathds{1}\otimes\bra{u})\ket\psi|^2\leq \|(\mathds{1}\otimes \bra{u})\ket\psi\|^2,
\end{equation}
implying that (using \eqref{varchar})
\begin{equation}
\label{varrew}
(\lambda^\psi_1)^2=\sup_{\substack{\ket\phi\in\otimes^{N-1}\mathbb{C}^d,\ \ket u\in\mathbb{C}^d\\\|\ket\phi\|=1,\ \|\ket u\|=1}}|\braket{\phi\otimes u|\psi}|^2=\sup_{\substack{\ket u\in\mathbb{C}^d\\\|\ket u\|=1}}\|(\mathds{1}\otimes \bra{u})\ket\psi\|^2.
\end{equation}
By 3. in Theorem \ref{recursion}, we can write 
\begin{equation}
\ket\psi=\sum_{t\in \Std(\nu)} P_t\ket\psi=\sum_{t\in \Std(\nu)} c_t \ket{v_t\otimes t},
\end{equation}
with $c_t:=\|P_t\ket\psi\|$, $\ket{v_t\otimes t}:=(c_t)^{-1}P_t\ket\psi$. So \eqref{varrew} becomes
\begin{equation}
\label{somre}
(\lambda^\psi_1)^2=\sup_{\substack{\ket u\in\mathbb{C}^d\\\|\ket u\|=1}}\|\sum_{t\in\Std(\nu)}c_t (\mathds{1}\otimes \bra{u})\ket{v_t\otimes t}\|^2.
\end{equation}
We now note that, for any $\ket u$, the vectors $(\mathds{1}\otimes \bra{u})\ket{v_t\otimes t}$ are orthogonal for different tableaux $t$. The reason is Corollary \ref{sympres}, together with the fact that $t\neq s$ implies $t^\downarrow\neq s^\downarrow$, and also the orthogonality from Theorem \ref{recursion}. This reduces \eqref{somre}, also using \eqref{varrew}, to
\begin{equation}
(\lambda^\psi_1)^2\leq\sum_{t\in\Std(\nu)} |c_t|^2\sup_{\substack{\ket u\in\mathbb{C}^d\\\|\ket u\|=1}}\| (\mathds{1}\otimes \bra{u})\ket{v_t\otimes t}\|^2=\sum_{t\in\Std(\nu)} |c_t|^2(\lambda^{v_t\otimes t}_1)^2\leq\max_{t\in\Std(\nu)}\sup_{\substack{\ket{v}\in V^\nu\\\|\ket{v}\|=1}}(\lambda^{v\otimes t}_1)^2.
\end{equation}
It proves the lemma since the opposite inequality is trivial. 
\end{proof}

The lemma above means we can restrict to vectors of the form $\ket{v\otimes t}$ with $t$ a standard tableau. But we do not have to consider all tableaux: it turns out that only the position of $N$ matters.

\begin{lemma}
\label{indep}
Let $\nu$ be a Young diagram of $N$ boxes and let $\ket{v}\in V^\nu$. Let $t\in\Std(\nu)$. The Schmidt coefficients $\lambda^{v\otimes t}_i$ of $\ket{\psi}=\ket{v\otimes t}$ in \eqref{thisSchmidt} depend only on $\ket{v}$ and the position of $N$ in $t$.
\end{lemma}
\begin{proof}
The Schmidt decomposition is
\begin{equation}
\ket{v\otimes t}=\sum^{d}_{i=1} \lambda^{v\otimes t}_i \ket{\phi_i} \otimes \ket{u_i}.
\end{equation}
By Corollary \ref{sympres}, there exist $\ket{v_i}\in V^{\nu(t^\downarrow)}$ such that $\ket{\phi_i}=\ket{v_i\otimes t^\downarrow}$, and $\braket{v_i|v_j}=\braket{\phi_i|\phi_j}=\delta_{ij}$. Now let $s\in\Std(\nu)$ have $N$ in the same position as $t$. Since $S^{\nu(t^\downarrow)}$ is irreducible, we can map $t^\downarrow$ to $s^\downarrow\in\Std(\nu(t^\downarrow))$ using linear combinations of permutations in $S_{N-1}$. But by Theorem \ref{Youngrule}, this also maps $t$ to $s$. The resulting Schmidt decomposition is
\begin{equation}
\ket{v\otimes s}=\sum^{d}_{i=1} \lambda^{v\otimes t}_i \ket{v_i\otimes s^\downarrow} \otimes \ket{u_i}.
\end{equation}
\end{proof}

We now know that we can simply consider the problem for tableaux $t$ with $N$ in the different removable boxes and the other entries put in at our convenience. For example, the diagram $\nu=(3,2,1)$ has three removable boxes
\begin{equation}
\begin{ytableau}
\phantom{.} &   &  \\
\phantom{.} &  \\
*(green) 6
\end{ytableau}\ \ \ \ \ \ \ 
\begin{ytableau}
  \phantom{.}&  & \\
  \phantom{.} & *(green) 6   \\
 \phantom{.}
\end{ytableau}
\ \ \ \ \ \ \ 
\begin{ytableau}
  \phantom{.}&  & *(green) 6\\
  \phantom{.} &   \\
 \phantom{.}
\end{ytableau}\ \ ,
\end{equation}
so it suffices to prove a bound for the tableaux with the remaining numbers in column-order, 
\begin{equation}
\label{someexamp}
\begin{ytableau}
 1 & 3  & 5 \\
 2 & 4 \\
*(green) 6
\end{ytableau}\ \ \ \ \ \ \ 
\begin{ytableau}
  1& 4 &  5\\
  2 & *(green) 6   \\
 3
\end{ytableau}
\ \ \ \ \ \ \ 
\begin{ytableau}
  1& 4 &  *(green)6\\
  2 & 5   \\
 3
\end{ytableau}\ \ .
\end{equation}
Corollary \ref{split} now allows us to further reduce these by tracing out the copies of $\mathbb{C}^d$ with numbers in the columns to the left of the box containing $N$ (we call this $\mathcal{H}_A$). The Schmidt vectors $\ket{\psi^i_B}$ of the relevant Schmidt decomposition \eqref{Schmidt} correspond to the  tableaux $t_B$,
\begin{equation}
\label{reddiag}
\begin{ytableau}
 1 & 3  & 5 \\
 2 & 4 \\
*(green) 6
\end{ytableau}\ \ \ \ \ \ \ 
\begin{ytableau}
 1 & 2\\
*(green) 3   
\end{ytableau}\ \ \ \ \ \ \ 
\begin{ytableau}
*(green) 1 
\end{ytableau}\ \ .
\end{equation}
That is, according to Corollary \ref{split}, the reduced density matrices $\rho_B:=\sum_{i}(\mu^\psi_i)^2\ket{\psi^i_B}\bra{\psi^i_B}$ with $\sum_{i}(\mu^\psi_i)^2=1$ have support on the spaces defined by these tableaux. By \eqref{varrew}, it then suffices to prove a bound for $t_B$ since
\begin{equation}
\label{ineq1}
\begin{aligned}
(\lambda^{v\otimes t}_1)^2&=\sup_{\substack{\ket{u}\in\mathbb{C}^d\\\|\ket{u}\|=1}}\|(\mathds{1}\otimes \bra{u})\ket{v\otimes t}\|^2\\
&=\sup_{\substack{\ket{u}\in\mathbb{C}^d\\\|\ket{u}\|=1}}\Tr_{AB}[(\mathds{1}\otimes\ketbra{u})\ketbra{v\otimes t}]\\
&=\sup_{\substack{\ket{u}\in\mathbb{C}^d\\\|\ket{u}\|=1}}\Tr_{B}[(\mathds{1}\otimes\ketbra{u})\rho_B]\\
&\leq\ \sum_{i}(\mu^\psi_i)^2(\lambda^{\psi^i_B}_1)^2\ \ \leq\sup_{\substack{\ket{v}\in V^{\nu(t_B)}\\\|\ket{v}\|=1}}(\lambda^{v\otimes t_B}_1)^2.
\end{aligned}
\end{equation}
The reader may now wonder why this would work: the maximization problem \eqref{result} suggests that a pure $\rho_B$ is the best choice, and we are assuming this happens---but why is that a reasonable thing to do? At least intuitively, the reason is that rows correspond to symmetrizations and columns to antisymmetrizations. This means that entanglement between rows cannot be avoided, but entanglement between columns can. We will see how this works in the next section.

Continuing with the proof, we have reduced the problem to tableaux like \eqref{reddiag}---these have the highest number in $(c_1,1)$, and are column-ordered otherwise. The upper bound now follows from the following lemma and the reformulation of the maximization problem from Theorem \ref{normlemma}. 

\begin{lemma}
\label{firstbox}
Let $\nu$ be a diagram of $N\geq 2$ boxes such that $(c_1,1)$ is removable and let $t\in\Std(\nu)$ be the tableau that has $N$ in box $(c_1,1)$ and is column-ordered otherwise. For any $\ket{v}\in V^\nu$, we have
\begin{equation}
(\lambda^{v\otimes t}_1)^2\leq \prod^{c_1-1}_{i=1}\left(1-\frac{1}{h_{(i,1)}}\right).
\end{equation}
\end{lemma}
\begin{proof}
Following Theorem \ref{normlemma}, let $\ket\phi\in\otimes^{N-1}\mathbb{C}^d$ and $\ket{u}\in\mathbb{C}^d$ be normalized. We start with two useful facts. First, by the recursion \eqref{defproj} from Theorem \ref{recursion}, we have
\begin{equation}
\label{firste}
\begin{aligned}
\|P_t(\ket\phi\otimes\ket{u})\|^2&=\bra{\phi\otimes u}P_t\ket{\phi\otimes u}\\
&=\frac{r_1!\dots r_{c_1}! c_1! \dots c_{r_1}!}{\Pi^{c_1}_{i=1}\Pi^{r_j}_{j=1}\ h_{(i,j)}}\bra{\phi\otimes u}P_{t^\downarrow}S_tA_tP_{t^\downarrow}\ket{\phi\otimes u},
\end{aligned}
\end{equation}
where we abbreviated $P_{t^\downarrow}\otimes\mathds{1}$ to $P_{t^\downarrow}$.
Second, by \eqref{colord} in Theorem \ref{azw},
\begin{equation}
\label{Ptdown}
P_{t^\downarrow}=\frac{r_1!\dots r_{c_1-1}! (c_1-1)! c_2! \dots c_{r_1}!}{(\Pi^{c_1}_{i=1}\Pi^{r_j}_{j=2}\ h_{(i,j)})\Pi^{c_1-1}_{i=1}(h_{(i,1)}-1)}A_{t^\downarrow}S_{t^\downarrow}A_{t^\downarrow}.
\end{equation}
Since $(c_1,1)$ is a removable box, we have $r_{c_1}=h_{(c_1,1)}=1$ and $S_t=S_{t^\downarrow}$. We also note that $A_t=A_{t^\downarrow}P^A_1$, where $P^A_1$ is the antisymmetrizer of the first column of $t$. And, by \eqref{Ptdown}, $P_{t^\downarrow}=P_{t^\downarrow}A_{t^\downarrow}$.
Plugging all these facts into \eqref{firste}, we obtain
\begin{equation}
\begin{aligned}
\|P_t(\ket\phi\otimes\ket{u})\|^2&=\frac{r_1!\dots r_{c_1}! c_1! \dots c_{r_1}!}{\Pi^{c_1}_{i=1}\Pi^{r_j}_{j=1}\ h_{(i,j)}}\bra{\phi\otimes u}P_{t^\downarrow}S_tA_tP_{t^\downarrow}\ket{\phi\otimes u}\\
&=\frac{r_1!\dots r_{c_1}! c_1! \dots c_{r_1}!}{\Pi^{c_1}_{i=1}\Pi^{r_j}_{j=1}\ h_{(i,j)}}\bra{\phi\otimes u}P_{t^\downarrow}A_{t^\downarrow}S_{t^\downarrow}A_{t^\downarrow}P^A_1P_{t^\downarrow}\ket{\phi\otimes u}\\
&=c_1\frac{\Pi^{c_1-1}_{i=1}(h_{(i,1)}-1)}{\Pi^{c_1}_{i=1}h_{(i,1)}}\bra{\phi\otimes u}(P_{t^\downarrow})^2P^A_1P_{t^\downarrow}\ket{\phi\otimes u}\\
&=c_1 \prod^{c_1-1}_{i=1}\left(1-\frac{1}{h_{(i,1)}}\right)\|P^A_1(P_{t^\downarrow}\ket\phi\otimes\ket{u})\|^2.
\end{aligned}
\end{equation}
Since $P^A_1$ only acts on the spaces $1,\dots,c_1-1$ and $N$, we can reduce $P_{t^\downarrow}\ket{\phi}$ to a density matrix on these spaces (cf.\ \eqref{ineq1}), and apply \eqref{estim} to its eigenvectors to conclude
\begin{equation}
\label{seconde}
\|P_t(\ket\phi\otimes\ket{u})\|^2\leq c_1 \prod^{c_1-1}_{i=1}\left(1-\frac{1}{h_{(i,1)}}\right)\frac{1}{c_1}\|P_{t^\downarrow}\ket\phi\|^2\leq\prod^{c_1-1}_{i=1}\left(1-\frac{1}{h_{(i,1)}}\right).
\end{equation}
\end{proof}

\subsection{Proof of sharpness of the bound}
\label{sharpn}
To work towards an optimizer, we first consider a special type of vector. These are known as coherent states, but in column-ordered tableaux they take a simple form in terms of Slater determinants \eqref{Slater} and that is all we need.
\begin{lemma}
\label{cohstate}
Let $t\in\Std_N$ be column-ordered. Let $\ket{u_1},\dots,\ket{u_{c_1}}\in\mathbb{C}^d$ be orthonormal. Then,
\begin{equation}
\label{coherentstate}
\ket{u_1\wedge\dots\wedge u_{c_1}}\otimes\ket{u_1\wedge\dots\wedge u_{c_2}}\otimes\dots\otimes\ket{u_1\wedge\dots\wedge u_{c_{r_1}}}\in \Im(P_t).
\end{equation}
\end{lemma}
\begin{proof}
We proceed by induction. The case $r_1=1$ gives a Slater determinant and this corresponds to a single column as claimed.
Now assume that we have proved the statement for $r_1-1$. Without loss of generality we can restrict to $d=c_1$ (this does not change how $P_t$ acts). Consider the first ordinary tensor product in \eqref{coherentstate}, that is, the one that separates spaces $1,\dots,c_1$ from the others. The Slater on the left of this cut again corresponds to a column diagram with $c_1$ boxes, which we denote $t_A$. By the induction assumption, the state to the right corresponds to a column-ordered tableau with $N-c_1$ boxes, which we denote $t_B$. We conclude that the vector lies in $\Im(P_{t_A})\otimes\Im(P_{t_B})$. By Corollary \ref{split}, $\Im(P_t)=V^\nu\otimes \ket{t}$ is contained in this space, but by the dimension formula \eqref{dimV}, these spaces have equal dimension for $d=c_1$. Hence, the spaces are equal and the state lies in $\Im(P_t)$.
\end{proof}

We now show that the individual bounds in \eqref{result} can be attained, implying that the maximum can also be. Before we start, it is good to see where the estimates in our upper bound were made. Lemmas \ref{firstlem} and \ref{indep} reduced the problem to certain vectors and do not provide estimates. The only inequalities appear in equations \eqref{ineq1} and \eqref{seconde}.

Now consider a diagram $\nu$. For it to appear in the Schur--Weyl decomposition \eqref{SchurWeyl}, we require the dimension $d\geq c_1$.

\begin{proposition}
Let $\nu$ be a diagram of $N$ boxes, $d\geq c_1$, and $\ket{u_1},\dots,\ket{u_{c_1}}\in\mathbb{C}^d$ orthonormal. Let $t$ be a tableau that has $N$ in a removable box $(c_l,l)$ and is column-ordered otherwise. As in Corollary \ref{split}, we use $t_A$ and $t_B$ to denote the tableaux resulting from a split across the vertical line to the left of the removable box, see \eqref{someexamp},\eqref{reddiag}. The state
\begin{equation}
\label{optimizer}
\begin{aligned}
\ket\psi:=&\underbrace{\vphantom{\frac{P_{t_B}(\ket\phi\otimes\ket{u_{c_l}})}{\|P_{t_B}(\ket\phi\otimes\ket{u_{c_l}})\|}}\ket{u_1\wedge\dots\wedge u_{c_1}}\otimes\dots\otimes\ket{u_1\wedge\dots\wedge u_{c_{l-1}}}}_{A}\\
&\hspace{0.7cm}\otimes\ \ \underbrace{\frac{P_{t_B}(\ket{u_1\wedge\dots\wedge u_{c_l-1}}\otimes \ket{u_1\wedge\dots\wedge u_{c_{l+1}}}\otimes\dots\otimes\ket{u_1\wedge\dots\wedge u_{c_{r_1}}}\otimes\ket{u_{c_l}})}{\|P_{t_B}(\ket{u_1\wedge\dots\wedge u_{c_l-1}}\otimes \ket{u_1\wedge\dots\wedge u_{c_{l+1}}}\otimes\dots\otimes\ket{u_1\wedge\dots\wedge u_{c_{r_1}}}\otimes\ket{u_{c_l}})\|}}_{B}
\end{aligned}
\end{equation}
saturates the bound in \eqref{result} (for the relevant removable box), and $\ket\psi\in\Im(P_t)$.
\end{proposition}
\begin{proof}
Since the statement is independent of dimension, we can restrict $\mathbb{C}^d$ to the subspace spanned by $\ket{u_1},\dots,\ket{u_{c_1}}$, and assume $d=c_1$ without loss of generality.

We first show that the bound is attained. Tracing out the $A$-part of $\ket\psi$ results in a pure state of the type considered in Lemma \ref{firstbox}, implying that the reduction step from \eqref{ineq1} is exact.

We are left with the $B$-part. Note that by Lemma \ref{cohstate},
\begin{equation}
\label{above}
\ket{u_1\wedge\dots\wedge u_{c_l-1}}\otimes \ket{u_1\wedge\dots\wedge u_{c_{l+1}}}\otimes\dots\otimes\ket{u_1\wedge\dots\wedge u_{c_{r_1}}} \in \Im(P_{(t_B)^\downarrow}),
\end{equation}
where $(t_B)^\downarrow$ is the column-ordered tableau with the final box removed. This implies that the last inequality in \eqref{seconde} is exact. By the condition for equality discussed below \eqref{estim}, the first is also exact since $\ket{u_{c_l}}$ is orthogonal to all the $\ket{u_i}$'s appearing in the Slater in \eqref{above}.

It remains to prove that $\ket\psi\in \Im(P_t)$. This follows from an argument similar to the one that was used to prove Lemma \ref{cohstate}: start from $t_B$ and work to the left by adding one Slater at a time, each time restricting the dimension to the size of column that is added.
\end{proof}

For completeness, we add two more remarks.

\begin{remarks}
\begin{enumerate}
\item Both \eqref{coherentstate} and \eqref{optimizer} correspond to special vectors in $\ket{v}\in V^\nu$ that are known as \textit{coherent states} or \textit{highest weight vectors}. Although we have not used this language in the paper, we mention that these are normally represented by semistandard tableaux with uniform rows, such as
\begin{equation}
\begin{ytableau}
 1 & 1  & 1 & 1 \\
 2 & 2 \\
3
\end{ytableau}\ \ .
\ytableausetup{boxsize=normal}
\end{equation}
Tensored with different $\ket{t}\in S^\nu$, these can look like \eqref{coherentstate} and \eqref{optimizer}, and of course all of $\ket{v}\otimes S^\nu$. 

\item The coherent states are not the only ones to satisfy the bound. It is easy to check with Theorem \ref{azw}, for example, that 
\begin{equation}
\frac{1}{\sqrt{2}}(\ket{u_1\wedge u_2} \otimes \ket{u_1\wedge u_3}+\ket{u_1\wedge u_3} \otimes \ket{u_1\wedge u_2})
\end{equation}
corresponds to the tableau
\begin{equation}
\begin{ytableau}
 1 & 3  \\
 2 & 4
\end{ytableau}\ ,
\end{equation}
and that it satifies the optimal bound 1/2. This is one of the reasons why it is not straightforward to give necessary and sufficient conditions for the maximum to be attained: the reduction $\rho_B$ in \eqref{ineq1} is not always a pure state. 
\end{enumerate}
\end{remarks}

\begin{acknowledgments}
This work was supported by the Royal Society through a Newton International Fellowship, by Darwin College Cambridge through a Schlumberger Research Fellowship, and, by membership of the HEP group in DAMTP, supported by STFC consolidated grant ST/P000681/1. I thank Graeme Mitchison for suggesting that I apply my algorithm to Young projections, causing me to formulate this problem. His enthusiasm is sorely missed.
\end{acknowledgments}


\begin{thebibliography}{99}

\bibitem{AlcockZeilingerWeigert}
Alcock-Zeilinger J., and Weigert H., ``Compact Hermitian Young projection operators,'' Journal of Mathematical Physics 58, 051702 (2017).

\bibitem{Klyachko}
Altunbulak M., and Klyachko A., ``The Pauli principle revisited,'' Communications in Mathematical Physics 282, 287--322 (2008).

\bibitem{BaconChuangHarrow}
Bacon D., Chuang I.L., and Harrow A., ``Efficient quantum circuits for Schur and Clebsch-Gordan transforms,'' Physical review letters 97, 170502 (2006).

\bibitem{Wolf}
Ba{\~n}uls M.-C., Cirac J.I., and Wolf M.M., ``Entanglement in fermionic systems,'' Physical Review A 76, 022311 (2007).

\bibitem{Cao}
Cao Z.-H., and Feng L.-H., ``A note on variational representation for singular values of matrix,'' Applied mathematics and computation 143, 559--563 (2003).

\bibitem{CLR}
Carlen E.A., Lieb E.H., and Reuvers R., ``Entropy and Entanglement Bounds for Reduced Density Matrices of Fermionic States,'' Communications in Mathematical Physics 344, 655--671 (2016).

\bibitem{recoupcoeff}
Christandl M., {\c{S}}ahino{\u{g}}lu M.B., and Walter M., ``Recoupling coefficients and quantum entropies,'' Annales Henri Poincar{\'e} 19, 385--410 (2018).

\bibitem{Coleman}
Coleman A.J., ``Structure of Fermion Density Matrices,'' Reviews of Modern Physics 35, 668--686 (1963).

\bibitem{Schoutens}
Haque M., Zozulya O.S., and Schoutens K., ``Entanglement between particle partitions in itinerant many-particle states,'' Journal of Physics A: Mathematical and Theoretical 42, 504012 (2009).

\bibitem{Hayashi}
Hayashi M., ``A Group Theoretic Approach to Quantum Information,'' Springer (2017).

\bibitem{GZ1}
Gelfand I.M., and Zetlin M.L., ``Finite-dimensional representations of the group of unimodular matrices,'' Dokl. Akad. Nauk SSSR 71, 825--828 (1950).

\bibitem{GZ2}
Gelfand I.M., and Zetlin M.L., ``Finite-dimensional representations of the group of orthogonal matrices,'' Dokl. Akad. Nauk SSSR 71, 1017--1020 (1950).

\bibitem{GoodmanNolan}
Goodman R., and Nolan R.W., ``Symmetry, representations, and invariants,'' Springer (2009).

\bibitem{Grudka}
Grudka A., Horodecki M., and Pankowski {\L}., ``Constructive counterexamples to the additivity of the minimum output R{\'e}nyi entropy of quantum channels for all $p>2$,'' Journal of Physics A: Mathematical and Theoretical 43, 425304 (2010).

\bibitem{KeppelerSjodahl}
Keppeler S., and Sj{\"o}dahl M., ``Hermitian Young operators,'' Journal of Mathematical Physics 55, 021702 (2014).

\bibitem{KeylWerner}
Keyl M., and Werner R.F., ``Estimating the spectrum of a density operator,'' Physical Review A 64, 052311 (2001).

\bibitem{HeisenbergYoung}
Nataf P., and Mila F., ``DMRG simulations of $SU(N)$ Heisenberg chains using standard Young tableaux: fundamental representation and comparison with finite-size Bethe ansatz,'' arXiv:1802.05482 (2018).

\bibitem{Okounkov}
Okounkov A., ``Young basis, Wick formula, and higher Capelli identities,'' arXiv:q-alg/9602027 (1996).

\bibitem{OkounkovVershik}
Okounkov A., and Vershik A., ``A new approach to representation theory of symmetric groups,'' Selecta Mathematica 2, 581--605 (1996).

\bibitem{RR}
Reuvers R., ``An algorithm to explore entanglement in small systems,'' Proceedings of the Royal Society A 474, 20180023 (2018).

\bibitem{entanglement2fermions}
Schliemann J., Cirac J.I., Ku{\'s} M., Lewenstein M., and Loss D., ``Quantum correlations in two-fermion systems,'' Physical Review A 64, 022303 (2001).

\bibitem{Thrall}
Thrall R.M., ``Young’s semi-normal representation of the symmetric group,'' Duke Mathematical Journal 8, 611--624 (1941).

\bibitem{Weyl}
Weyl H., ``The theory of groups and quantum mechanics,'' Courier Corporation (1950).

\bibitem{Young}
Young A., ``On quantitative substitutional analysis (sixth paper),'' Proceedings of the London Mathematical Society s2-34, 196--230 (1932).


\end{thebibliography}
\end{document}